\documentclass[aps,preprint]{revtex4}
\usepackage{amsfonts}
\usepackage{amsmath}
\usepackage{amssymb}
\usepackage{graphicx}

\input{tcilatex}
\begin{document}

\title{On effective radiational gravity acceleration at the interface of
dense plasmas and vacuum}
\author{$^{1}$A. Asma, $^{2}$Ch. Rzoina ,$^{3}$I. Zafar and $^{4,5}$S. Poedts%
}
\affiliation{Department of Physics, FC College (A Chartered University), Lahore 54000,
Pakistan}
\affiliation{$^{2}$Department of Physics, GGC for Women. Gulberg, Lahore, Pakistan}
\affiliation{$^{3}$Salam Chair, Department of Physics, G. C. University Lahore, Katchery
Road, Lahore 54000, Pakistan}
\affiliation{$^{4}$Centre for Mathematical Plasma Astrophysics, Department of
Mathematics, CmPA, KU Leuven, Celestijnenlaan 200B, 3001 Leuven, Belgium:
rozina.chaudhary@student.kuleuven.be: \ drchrozina@gmail.com \ }
\affiliation{$^{5}$Institute of Physics, University of Maria Curie-Sklodowska, Pl. M.
Curie-Sklodowska 5, 20-031 Lublin, Poland}

\begin{abstract}
\ Radiative, gravitational surface waves are investigated at the interface
of high density and low temperature, magnetized \ incompressible electron-
ion plasmas in the presence of dense radiation electromagnetic radiation
pressure (DEMRP). The inhomogeneous \textbf{embedded} is reported \ at
plasma-vacuum interfcae.The DEMRP is found to stabilize the surface waves,
however for a specific case, it tends to enhance the \ growth rate of
surface waves via the frictional instability.The group velocity of
gravitational radiation is shown to be the function of wavelength. The
obtained analytical results are presented both numerically and graphically
as function of DEMRP. to show that the incorporation of \ DEMRP may
introduce quite different dispersive properties of charged surface wave
phenomena. It is shown numerically that the frequency of the obtained
radiative gravitational waves in the presence of DEMRP is found to lie in\
the range of high frequency radio waves, while in case of rare laboratory
plasma, the frequency of these waves is found to lie\ within the very low
frequency radio waves of the electromagnetic radiation spectrum.This work
may enhance the gravitational aspects of electromagnetic radiations in dense
\ astrophysical systems such as neutron star and white dwarfs.
\end{abstract}

\maketitle

\section{Introduction}

Electromagnetic radiation pressure (EMRP), being an operative tool to
transmit momentum\ to plasma particles, has engrossed a lot of attention 
\cite{1}. EMRP is applicable to various physical situations extending from
particle acceleration in the laboratory \cite{5,6} to astrophysical
environments \cite{7}, radiation generated winds \cite{2,3}, photon bubbles
stars and accretion disks \cite{4} etc. Generally speaking EMRP is generated
by the interaction of the radiation with medium such as an electron-ion
plasma, as a result the ions are accelerated by the charge separation field
generated by electrons, which are being pushed by both thermal and radiation
pressures. The  {existence} of RP (Ponderomotive force) in various
systems has been reported experimentally\cite{8,9,10}, EMRP is of practical
interest to drive high energy ions via laser source \cite{11}. In this
context, while considering the particle nature of photon, the investigation
of novel features of RP of electromagnetic (thermal) radiations at the
plasma--vacuum interface is an important research area.

In a plasma electron performs a quantum jump in the electric field of an ion
(or a positron), consequently it may emit or may engross a photon, this
oscillatory motions of electrons may disturb the radiation field  {and
may }change the dispersion law of the radiation field. Eventually, the
thermal radiation may introduce the new thermodynamics, provided the thermal
radiations are considered in a dense plasma medium \cite{lev}. In this
scenario, a detailed modified thermodynamics was presented in \cite{lev}, to
show that the REMRP of photon gas in a rare plasma $(\hbar \omega _{ps}\ll
k_{B}T_{s})$ may read as\textbf{\ }$P_{_{s,r}}^{\gamma }=\frac{U_{_{s,r}}^{r}%
}{3}=\frac{\alpha _{r}\text{ }(k_{B}T_{s})^{4}}{3}$(where $s=e$ or $i$,
denotes the electrons or ions respectively) $\alpha _{r}=\pi ^{2}/45(\hbar
c)^{3}$ is the Stefann Boltzmann constant and $\omega _{ps}^{2}=(\frac{4\pi
n_{0s}e^{2}}{m_{s}})$ is the electron plasma frequency, whereas the DEMRP $%
(P_{_{s,d\text{ }}}^{r})$, in case of high density and low temperature
plasma ($\hbar \omega _{ps}>>k_{B}T_{s}$), is%
\begin{equation}
P_{_{s,d\text{ }}}^{\gamma }=\frac{U_{_{sd}}^{r}}{3}=\beta
_{d,s}T_{d,s}^{5/2},  \tag{1}
\end{equation}%
here $\beta _{d,s}$ is the radiation coefficient defined as $\left( \beta
_{d,s}=0.202(\hbar \text{ }\omega _{ps})^{3/2}/(\hbar c)^{3}\right) $ with
all standard notations and temperature in energy units. The impact of DEMRP 
{shown in Eq.\ (1) may} significantly alter various plasma dispersive
modes \cite{n5,n6}.

On the other hand gravity is an important force of nature which plays a
vital role in most astronomical systems. However, the gravitational force
has smaller transmission capacity as compare to light, so gravitational
ripples known as gravitational radiations may reach to other objects within
finite time. Presently the gravitational radiation is one of the last un
opened and important windows into the Universe, i.e., if someone picks up
book from table, means that person has overcome the collective gravitational
pull of planet Earth on the book, this energy comes from the muscular
chemical force of that person, arising from electromagnetic interaction \cite%
{gr}. In short gravitational and electromagnetic forces are two important
forces of nature, specifically their oscillations at the interface between
two plasma medium or vacuum plasma interfaces is an important research area
to understand various natural and laboratory phenomena.

The propagation of surface waves (SWs) has attracted considerable attention
due to its relevance with real experimental and astrophysical plasma
dynamics \cite{n1,n2}. One of the important types of surface waves includes
the inhomogeniety of the fluid caused by the separation of charges,  {%
due to thermal effects}, as function of the surface coordinates\cite{n1,n3}.
Such inhomogeneous waves may propagate on the interface of an incompressible%
 {\ fluid in a fashion} that pressure varies with height. The
consequent motion of electrons in height may disturb the equilibrium state,
resulting to an oscillatory motion on the plasma -vacuum interface, known as
magneto, capillary, gravity waves, for example when two clouds collides the
scintillations can be seen on \ the cloud interface. The electromagnetic
energy of the SWs is confined to the regions of inhomogeneous plasmas \cite%
{14}. The stability analysis of such charged, gravity surface waves at the
interface of an incompressible plasma and vacuum \ have been investigated by
several authors \cite{n1,n4,n5,n6}. It was shown in these papers that single
fluid magneto-hydrodynamic oscillations does not hold good in the presence
of larger thermal speed of electrons as compare to the ions because, the
electrons moves towards the plasma- vacuum interface more frequently and may
get diffuse there. In other words potential field on the surface is
maintained by the thermal motion of the electrons \cite{n1} such that, $%
e\phi \sim $ $k_{B}T_{e}$.

Taking into account all previous works mentioned above, presently we aim to
investigate the impact of DEMRP shown in  {Eqs. (1) on the stability
analysis} of gravitational waves at the interface of classical, dense
electron ion plasma and vacuum.  {To address some new features of
gravitational radiations, we shall follow the method adopted in \cite{n4} to
show that inhomogeneous- gravitational radiation as a function of electron
temperature may appear in the transitional region of} vacuum-plasma
interface, which under suitable conditions, may account for both stability
and instability features of charged surface waves.  {We aim to show
numerically, that DEMRP shown in Eq. (1)\ and REMRP presented in Eq. (1.3)
of ref.\ \cite{n4}, belongs to different gravitational radiation spectrum of
thermal radiations.}

\section{Basic model}

 {\ It was mentioned by Landau and Lifshitz \cite{n3} that in many
cases of flow of gases, their density may be supposed invariable i.e.,
constant throughout the volume of the gas and throughout its motion.In other
words, there is no noticeable compression or expansion of the fluid in such
cases, hence such fluids can be named as incompressible. In this context,
the stability analysis of charged surface waves, at the interface of
incompressible plasma and vacuum, is issued in various works \cite{n1,n5,s1}.%
}

 {In order to express our contribution to the already existing
literature on the charge surface waves, we use the one fluid MHD model
presented in \cite{n4} for the plasma with cosnstant density.We consider an
incompressible elrctron- ion plasma in the presence of DEMRP, magnetic and
gravitational forces, embedded by an external magnetic field }$B=B_{0}\hat{z}
$ {\ directed along the }$z-axis,$ {\ (}$\hat{z}$ {\ is
the unit vector along }$z-axis$ {).} We assume plasma under
consideration to be in the $y-z$ plane and normal to the plane is $x-$%
direction, which is assumed to be vacuum. We further consider a uniform
temperature inside the plasma, but  small temperature jumps may appear in
the transitional region of plasma- vacuum interface.  {It is worth to
mention that since we have considered an incompressible plasma i.e., the
density of both electrons and ions is constant in space and time, so the
quantum effects are not incorporated in present analysis. }For our
investigations one fluid MHD model may be achieved by ignoring the inertia
of electrons, while taking into account ions inertia i.e.,

\begin{equation}
en_{e}\mathbf{E}-\frac{e}{c}n_{e}\left( \mathbf{v}_{e}\times \mathbf{B}%
\right) -\mathbf{\nabla }\left( P_{g}+P_{d,e}^{\gamma }\right) =0,  \tag{2}
\end{equation}%
\ \ \ 
\begin{equation}
m_{i}n_{i}\frac{d\mathbf{v}_{i}}{dt}=en_{i}\mathbf{E+}\frac{en_{i}}{c}(%
\mathbf{v}_{i}\times \mathbf{B)-}m_{i}n_{i}\mathbf{g},  \tag{3}
\end{equation}%
where $n_{e}$ and $n_{i}$ are the number density of electron and ion fluid, $%
\mathbf{v}_{e}$ and $\mathbf{v}_{i}$ are their respective speeds, $P_{g}$ is
the usual gas pressure for electron- ions, and $P_{d,e}^{\gamma }$ is the
DEMRP of electron fluid shown in Eq. (1),  {which is valid only for
the plasmas dense enough such that their plasma energy (}$\hbar \omega _{ps}$%
) {\ is more than the thermal energy (}$k_{B}T_{s}$ {) i.e., }$%
\hbar \omega _{ps}>>k_{B}T_{s}$ {. Consequently, the ion density (}$%
n_{e}=n_{i}=n_{0}$ {) along with their mass play vital role to enhance
the ion gravitational effects, shown in Eq. (3). In this scenario, lots of
works are reported in the literature, where the ion gravitational effects
have been taken into account \cite{chen, Ali,rozina,n1,n5}}.

 {It is well known that the one-fluid approximation assumes that
charge particles are rigidly bound to each other by electric forces, and
that the plasma is electrically neutral at each point. However, this
assumption is not strictly fulfilled in the presence of a large gradient in
the charge density\cite{n1}, in particular on the surface between the plasma
and the vacuum. Since electrons are lighter than ions, on the surface
conditions are favorable for the diffusion of the electron gas with respect
to the ions. Therefore, surface space charges appear in the transition
region i.e., the polarization of the surface is accompanied by the creation
of a strong electric field \cite{n1}.This field prevents further
polarization and inhibits the diffusion electron current.In the absence of
external fields, the separation of charge and the potential field in the
transition region are maintained only by the thermal motion of the electrons
i.e., }$e\delta \phi \sim k_{B}\delta T_{e}$ {. Consequently, the
electron surface charge will be normal to the interface and follow the
Poisson's law in terms of surface charge density as \cite{n1}},

\begin{equation}
\mathbf{\nabla.E}=-\nabla^{2}\phi=4\pi\sigma_{e},  \tag{4}
\end{equation}
here $\sigma_{e}=en_{e}$ is the electron charge density. Since plasma under
consideration is incompressible, we assumed that density perturbations are
not significant (incompressibility assumption).The continuity equation (mass
conservation) of both fluids are as follows:

\begin{equation}
\frac{\partial n_{e,i}}{\partial t}+\nabla.(n_{e,i}\mathbf{u}_{e,i})=0, 
\tag{5}
\end{equation}
Maxwell's equations%
\begin{equation}
\mathbf{\nabla.B=0,}  \tag{6}
\end{equation}%
\begin{equation}
\mathbf{\nabla}\times\mathbf{B}=\frac{4\pi}{c}\mathbf{J,}  \tag{7}
\end{equation}%
\begin{equation}
\mathbf{\nabla}\times\mathbf{E=-}\frac{\partial\mathbf{B}}{\partial t} 
\tag{8}
\end{equation}
and the MHD equation

\begin{equation}
\mathbf{E}+\mathbf{v}\times\mathbf{B=}\bar{\eta}\mathbf{J}  \tag{9}
\end{equation}
while $\mathbf{J}=-e\left( -n_{i}\mathbf{v}_{i}+n_{e}\mathbf{v}_{e}\right) $
is the charge current density and $\bar{\eta}$ is the thermal conductivity
of the material, where $\mathbf{v}=(m_{e}\mathbf{v}_{e}+m_{i}\mathbf{v}%
_{i})/(m_{e}+m_{i})$ is the electron effective mass. \ Adding Eqs. (2) and
(3)

\begin{equation}
\mathbf{\rho}_{i}\frac{d\mathbf{v}_{i}}{dt}=e(n_{i}-n_{e})\text{ }\mathbf{E+}%
\frac{1}{c}(\mathbf{J\times B)-\nabla}\left( P_{g}+P_{d}^{\gamma }\right)
-\rho_{i}\mathbf{g},  \tag{10}
\end{equation}
Eq. (4) can be used to calculate the associated negative pressure gradient
acting normal to the interface as

\begin{equation}
e(n_{i}-n_{e})\text{ }\mathbf{E=\sigma}_{e}\mathbf{E\mathbf{=}}\frac{1}{4\pi 
}\mathbf{E}\text{ }(\nabla.\text{ }\mathbf{\mathbf{E)}=}\frac{1}{8\pi}%
\nabla_{x}\mathbf{E}^{2}  \tag{11}
\end{equation}
where $\nabla_{x}=\frac{\partial}{\partial x}$ and $\mathbf{E}=-\nabla
\phi=-(\frac{\partial\phi}{\partial x})$. Coupling Eqs.(8-11), we eventually
arrive at

\begin{equation}
\rho_{i}\frac{d\mathbf{v}_{i}}{dt}=\frac{1}{8\pi}\mathbf{\nabla}E^{2}+\frac {%
1}{4\pi}\mathbf{B}\left( \mathbf{B.\nabla}\right) \mathbf{-\nabla}\left(
P_{g}+P_{d}^{\gamma}+\frac{B^{2}}{8\pi}\right) -\rho_{i}\mathbf{g}  \tag{12}
\end{equation}
where $\rho_{i}=m_{i}n_{i0}$ is the ion mass density and the first term on
the r.h.s. represents the negative pressure gradient, which acts normal to
the plasma-vacuum interface. Since the plasma under consideration is
incompressible, so the continuity equation becomes $\mathbf{\nabla.v}=0$,
which leads to another definition of \ fluid velocity i.e., $\mathbf{%
v=\nabla }\Psi$ \cite{n4},so incompressibility equation further reduces to
where $\Psi$ is the gravitational potential. The equilibrium electric field $%
E_{0}$, may be calculated from Eq. (4)

\begin{equation}
E_{0}\text{ }\mathbf{=}\text{ }4\pi\int{}\sigma_{e}dx\text{ }\mathbf{=}\text{
}4\pi\sigma_{se},  \tag{13}
\end{equation}
where $\sigma_{se}=\sigma_{e}x$\ \ denotes the equilibrium electrons surface
charge density and $x$ is the surface coordinate, the associated
electrostatic potential, $\phi$ is

\begin{equation}
\phi=-4\pi\sigma_{se}x  \tag{14}
\end{equation}
Next, if the interface undergoes small perturbations such that the surface
coordinate $x$ undergoes small surface displacement $\zeta\left(
y,z,t\right) $ above the equilibrium, then the associated oscillating
electrostatic potential $\delta\phi$ can be deduced from Eq. (14) as

\begin{equation}
\delta\phi=4\pi\sigma_{se}\text{ }\zeta\left( y,z,t\right) ,  \tag{15}
\end{equation}
We consider here small amplitude surface waves, which describes the minor
differences in the potential functions as, $\delta\phi\symbol{126}\exp\left[
i\left( k_{y}y+k_{z}z-\omega t\right) -kx\right] $, where $k>0$ is a
positive number such that $k^{2}=\left( k_{y}^{2}+k_{z}^{2}\right) $. It
maybe noted here that in our present consideration $\delta\phi$ \ decays
exponentially in the $x-$direction and eventually may disappear for the $%
x\rightarrow\infty$. We can write from Eq.'s (13-15)

\begin{equation}
\frac{E_{x}^{2}}{8\pi }=2\pi \sigma _{se}^{2}+k\sigma _{se}\delta \phi
|_{x=0}=2\pi \sigma _{se}^{2}+4\pi \sigma _{se}^{2}k\zeta \left(
y,z,t\right) ,  \tag{16}
\end{equation}%
where $\delta E_{x}=$ $k\delta \phi $. In linear approximation the
perpendicular component of the ion velocity $v_{i,x}$ is defined as $%
v_{i,x}=\partial \zeta \left( y,z,t\right) /\partial t.$ Since for an
incompressible fluid the velocity can be expressed as gradient of scalar
function, $v_{i,x}=\partial \Psi /\partial x,$ so one may equate the two
definitions to obtain $\partial \Psi /\partial x|_{x=0}=\partial \zeta
/\partial t$.  {Since we have considered an incompressible electron
ion plasma, so the usual definition of thermal pressure will not work here,
instead the plasma under consideration may obey the Laplace's formula \cite%
{n3}}

\begin{equation}
P-P_{0}=-\epsilon\left( \frac{\partial^{2}\zeta}{\partial y^{2}}+\frac{%
\partial^{2}\zeta}{\partial z^{2}}\right) .  \tag{17}
\end{equation}
where $P$ and $P_{0}$ are the pressure terms attributable to the two
different mediums (plasma and vacuum, respectively), and $\epsilon$ is the
surface tension coefficient. Next by following the method mentioned in \cite%
{n6}, we may get an evolution equation of electrons for a sharp
plasma-vacuum boundary

\begin{equation}
\left[ \frac{\partial^{2}\Psi}{\partial t^{2}}-V_{E}^{2}\text{ }k\frac{%
\partial\Psi}{\partial x}-V_{A}^{2}\frac{\partial^{2}\Psi}{\partial z^{2}}+g%
\frac{\partial\Psi}{\partial x}+\frac{1}{\rho_{i}}\left( \frac{\partial
P_{d}^{\gamma}}{\partial T_{d}}\right) \frac{\partial\delta T_{d}}{\partial t%
}-\frac{\epsilon}{\rho_{i}}\left( \frac{\partial^{2}}{\partial y^{2}}+\frac{%
\partial^{2}}{\partial z^{2}}\right) \frac {\partial\Psi}{\partial x}\right]
|_{x=0}=0.  \tag{18}
\end{equation}
Here $\delta T_{d}$, signifies small dense electron temperature jumps on the
interface, $V_{A}=B_{\mathbf{0}}/\sqrt{4\pi\rho_{i}})$ and $V_{E}=(E_{0}/%
\sqrt{4\pi\rho_{i}})$ are the magnetic and electric Alfven speeds of ions
respectively. Next, in order to express $\delta T_{d}$ in terms of scalar
function $\Psi$, we shall make use the energy equation for electrons.

\begin{equation}
\frac{\partial}{\partial t}\left( \frac{3}{2}k_{B}T_{d}+\frac{U_{_{sd}}^{r}}{%
n}\right) +\left( \mathbf{v}\cdot\mathbf{\nabla}\right) \left[ (\frac {3}{2}%
k_{B}T_{d}+\frac{P_{g}}{n})+\frac{4}{3}\frac{U_{_{sd}}^{r}}{n}\right] +%
\mathbf{\nabla}\cdot\frac{\mathbf{S}}{n}=0,  \tag{19}
\end{equation}
here $T_{d}$ and $n$ are the (non-relativistic) dense temperature and
density at quasi-equilibrium, expressed as $n$ $(\simeq
n_{e,0}=n_{i,0}),U_{_{sd}}^{r}$ is the electron radiation energy density
defined in Eq. (1) and $P_{g}$ is the usual gas pressure. The electron
radiative heat conduction can be determined by the Poynting vector $\mathbf{S%
}$ as

\begin{equation}
\mathbf{S}\text{ }\mathbf{=-}\frac{\lambda c}{3}\mathbf{\nabla}U=\mathbf{-}%
L_{0}\mathbf{\nabla(}k_{B}T_{d})  \tag{20}
\end{equation}
here $L_{0}=5/2\left( \lambda c/3\right) \beta_{d,e}\left( k_{B}T_{d}\right)
^{\frac{3}{2}}$ ) represents the coefficient of thermal radiation
conductivity for dense plasma, $\lambda=AT^{\kappa}$ is the Rosseland
radiation mean free path, where $\kappa=1,2,3,..$.( a positive integer), $A$
is a positive quantity and $k_{B}$ is Boltzmann's constant. We may re-write
Eq. (20)

\begin{equation}
\mathbf{\nabla}.\mathbf{S}\text{ }\mathbf{=}-\frac{5\beta\text{ }c\text{ }A}{%
6}\frac{1}{\kappa}\mathbf{\nabla}.\left( \mathbf{\nabla}T_{d}^{\kappa
}\right)  \tag{21}
\end{equation}
substituting $T_{d}=T_{0d}+\delta T_{d}$ 
\begin{equation}
\mathbf{\nabla}.\mathbf{S=}\mathbf{-}\frac{5\beta_{d}\text{ }cA}{6}(2\mathbf{%
\nabla}T_{0d}^{\kappa-1}\mathbf{\nabla}\delta T_{d}+\Delta
T_{0d}^{\kappa-1}\delta T_{d})  \tag{22}
\end{equation}
where $T_{0d}$ specifies the dense temperature at equilibrium, which is a
function of the surface coordinates and $\delta T_{d}$ denotes the small
temperature jumps on the dense plasma interface, applying Fourier
transformation $viz$ $\delta\phi\symbol{126}\exp\left[ i\left(
k_{y}y+k_{z}z-\omega t\right) +kx\right] $, we may obtain

\begin{equation}
\mathbf{\nabla}.\frac{\mathbf{S}}{n}\text{ }\mathbf{=-}\frac{5\beta _{d,e}%
\text{ }cA}{6n}\left( 2k\frac{dT_{0d}^{\kappa-1}}{dx}+\frac{%
d^{2}T_{0d}^{\kappa-1}}{dx^{2}}\right) \delta T_{d},  \tag{23}
\end{equation}
At equilibrium, Eq. (19) reduces to $T_{0d}=\theta|x_{s}-x|^{1/\kappa}$\ to
depict that even the equilibrium temperature$T_{0d\text{ }}$is a function of
surface coordinate $x,$ with $\theta$ as a constant temperature inside
plasma. It also demonstrates that at $x=x_{s}$, the temperature decreases,
implying the sharp boundary across the plasma-vacuum interface. Substituting
Eq. (23) into the linearized $x-$ dimensional form of Eq. (19) may lead to

\begin{equation}
\left[ 
\begin{array}{c}
\frac{\partial}{\partial t}\left( \frac{3}{2}+\frac{1}{n}\frac{U_{_{sd}}^{r}%
}{\partial T_{0d}}\right) \delta T_{d}+\frac{\partial\Psi}{\partial x}\frac{%
\partial}{\partial x}\left( T_{0d}+\frac{P_{g}}{n}+\frac{4}{3}\frac{%
U_{_{sd}}^{r}}{n}\right) \\ 
\mathbf{-}\frac{5\beta_{d}\text{ }c\text{ }A}{6n}\left( 2k\frac {%
dT_{0d}^{\kappa-1}}{dx}+\frac{d^{2}T_{0d}^{\kappa-1}}{dx^{2}}\right) \text{ }%
\}\delta T_{d}%
\end{array}
\right] |_{=0}=0,  \tag{24}
\end{equation}
Remember that the density $n$ $(\simeq n_{e,0}=n_{i,0})$\ is constant in
time and space invariant. We may re-write above equation as%
\begin{equation}
\delta T_{d}\mathbf{=}\frac{\mu_{d}}{\digamma_{d}\left[ i\omega+\nu _{d}%
\right] }  \tag{25}
\end{equation}
where 
\begin{align}
\mu_{d} & =\frac{\partial\Psi}{\partial x}\frac{\partial}{\partial x}\left( 
\frac{3}{2}T_{0d}+\frac{P_{g}}{n}+\frac{4}{3}\frac{U_{_{sd}}^{r}}{n}\right) 
\notag \\
\text{ }\digamma_{d} & =\frac{3}{2}+\frac{1}{n}\frac{\partial U_{_{sd}}^{r}}{%
\partial T_{0d}}  \notag \\
\text{ }\nu_{d} & =\frac{5\beta_{d}\text{ }c\text{ }A}{6n\left( \frac{3}{2}+%
\frac{1}{n}\frac{\partial U}{\partial T_{0d}}\right) }\left( 2k\frac{%
dT_{0d}^{\kappa-1}}{dx}+\frac{d^{2}T_{0d}^{\kappa-1}}{dx^{2}}\right) 
\tag{26}
\end{align}
By substituting Eq. (25) into Eq. (18), we eventually arrive at

\begin{equation}
\omega^{2}=k^{2}(V_{A}^{2}\cos^{2}\theta-V_{E}^{2})+\frac{b^{2}k^{3}}{2}%
+kg_{d}^{\gamma}+k\left( -ig_{d}^{\gamma^{\prime}}\frac{\left( \nu
_{d}/\omega\right) }{1+\left( \nu_{d}^{2}/\omega^{2}\right) }\right) +kg 
\tag{27}
\end{equation}
where $b^{2}(=2\epsilon/\rho_{i\text{ }})$ is the the capillarity constant,
also 
\begin{equation}
g_{d}^{\gamma}=\frac{5\beta_{d,e}T_{0d}^{3/2}}{6\rho_{i}}\frac{\left( \frac{3%
}{2}+\frac{1}{n}\frac{\partial P}{\partial T_{0d}}+\frac{4}{3n}\frac{%
\partial U}{\partial T_{0d}}\right) \frac{\partial T_{0d}}{\partial x}}{%
\left( \frac{3}{2}+\frac{1}{n}\frac{\partial U_{d}^{\gamma}}{\partial T_{0d}}%
\right) \left( 1+\frac{\nu_{d}^{2}}{\omega^{2}}\right) }\text{ \ }  \tag{28}
\end{equation}
and%
\begin{equation}
g_{d}^{\gamma^{\prime}}=\frac{5\beta_{d,e}T_{0d}^{3/2}}{6\rho_{i}}\frac{%
\left( \frac{3}{2}+\frac{1}{n}\frac{\partial P}{\partial T_{0d}}+\frac{4}{3n}%
\frac{\partial U}{\partial T_{0d}}\right) \frac{\partial T_{0d}}{\partial x}%
}{\left( \frac{3}{2}+\frac{1}{n}\frac{\partial U}{\partial T_{0d}}\right) } 
\tag{29}
\end{equation}

are the\ radiative gravity coefficients of dense plasma.\qquad\qquad\ \ \ \
\ \ \ \ \ \ \ \ \ \ \ \ \ \ \ \ \ \ \ \ \ \ \ \ \ 

\textbf{Radiative- gravitational waves}

In order to formulate a dispersion relation of charged surface magneto,
radiative, gravitational waves, initially we assume that the radiative heat
flux appearing $S$ is absent in (19) and hence in (27) $(viz.S=0)$ to get

\begin{equation}
\omega^{2}=k^{2}V_{A}^{2}\cos^{2}\theta+\frac{b^{2}k^{3}}{2}%
+k(g+g_{d}^{\gamma})-k^{2}V_{E}^{2},  \tag{30}
\end{equation}
above dispersion relations admits that the surface charge instability
criteria is function of surface charge density via $V_{E}$ with growth rate

\begin{equation}
\func{Im}\omega=\sqrt{k^{2}V_{E}^{2}-k(g+g_{d}^{\gamma}+kV_{A}^{2}\cos^{2}%
\theta+\frac{b^{2}k^{2}}{2}),}  \tag{31}
\end{equation}
\ \ \ \ \ \ \ \ \ \ \ \ \ \ \ \ \ \ \ \ \ \ \ \ \ \ \ \ \ \ \ \ \ \ \ \ \ \
\ Here the radiative gravitational acceleration ($g_{d}^{\gamma}$),
gravitational acceleration ($g$), magnetic field ($V_{A}$) and surface
tension ($b$) are coupled together to reduce the oscillating frequency of
charged surface waves. In order to see the impact of pure radiative gravity
waves on the surface, we just ignore the surface tension and magnetic field
in Eq. (31) to get

\begin{equation}
\omega ^{2}=k(g+g_{d}^{\gamma })-k^{2}V_{E}^{2},  \tag{32}
\end{equation}%
One can see from above dispersion equation that the radiation does not
change the spatial dispersion relation of longitudinal waves, instead it
appears only as gravitational radiation on the plasma vacuum surface.
Furthermore, in the absence of surface charge and gravitational acceleration 
$g$, pure radiative gravity waves may propagate on the interface as function
of wave number $k$ $(\lambda =2\pi /k)$ as

\begin{equation}
\omega^{2}=kg_{d}^{\gamma}  \tag{33}
\end{equation}
the associated group velocity of gravitational radiation of dense plasma is%
\begin{equation}
\frac{d\omega}{dk}=V_{gd}=\sqrt{\frac{g_{d}^{\gamma}}{4k}}=\sqrt{\frac {%
g_{d}^{\gamma}}{8\pi}\lambda,}  \tag{34}
\end{equation}
\ \ \ \ \ \ \ \ \ \ \ \ \ \ \ \ \ \ \ \ \ \ \ \ \ \ \ \ \ \ \ \ \ \ \ \ \ \
\ \ \ \ \ \ \ \ \ \ \ \ \ 

\textbf{Frictional instability}

 {Here, we shall show that the astrophysical objects may undergo
surface oscillations(fragmentation) through the surface instabilities as
function of frictional or dissipative, radiative gravitational acceleration.
For our purposes, we assume that if } $\nu _{d}\neq 0$ i.e.,  {in the
presence of electron radiative heat conduction appearing through the
Poynting vector in Eq. (27) via }$v_{d}$ and $g_{d}^{\gamma ^{\prime }}$%
 {,} there exist some specific values of propagation vector

\begin{equation}
k_{\pm }=(1/gb^{2})(V_{E}^{2}-V_{A}^{2}\cos ^{2}\theta )\pm (1/gb^{2})\sqrt{%
(V_{E}^{2}-V_{A}^{2}\cos ^{2}\theta )^{2}-(gb^{2})^{2}(g_{d}^{\gamma
^{\prime }}+g)},  \tag{35}
\end{equation}%
for which (27) reads as%
\begin{equation}
\omega ^{2}+\frac{ig_{d}^{\gamma ^{\prime }}(v_{d}/\omega )}{%
1+(v_{d}^{2}/\omega ^{2})}=0,  \tag{36}
\end{equation}%
 {provided}  {the imaginary term is more than other terms on
r.h.s of Eq. (27). }Further by assuming that $|v_{d}^{2}|<<\omega ^{2}$,
i.e, radiation acts for a short time, the above relation reduces to

\begin{equation}
\omega ^{3}+ik_{\pm }\text{ }g_{d}^{\gamma ^{\prime }}|v_{d}|\approx 0, 
\tag{37}
\end{equation}%
One of the three possible complex roots of the above equation leads to the
frictional instability of surface of a astrophysical objects as function of
electromagnetic \ (thermal) radiation via $v_{d}$ and $g_{d}^{\gamma
^{\prime }}$with respective growth rate%
\begin{equation}
\hat{\gamma}_{d}=\frac{\sqrt{3}}{2}k_{\pm }\text{ }g_{d}^{\gamma ^{\prime
}}|v_{d}|^{1/3},  \tag{38}
\end{equation}%
 {Equation (38) demonstrates that the growth rate of the radiative
dissipative instability,}$\hat{\gamma}_{d},$ {\ becomes function of
inhomogeneous radiative, gravitational acceleration} $(g_{d}^{\gamma
^{\prime }})$ {\ and }$v_{d}$  {shown in Eqs. (26,29).}

\section{Numerical Analysis}

 {We have used (31, 33, 34, 38) to analyze both numerically and
graphically, the dispersion relation as well as the growth rate of unstable
modes obtained for the dense plasma under study.\ It is well known from the
literature that magnetized compact objects like neutron star (NS) may
produce strong electromagnetic (EM) waves: pulsars (strongly magnetized NS)
produce bright radio emission, while magnetars may generate fast radio
bursts \cite{a2,a3}.\ Except\ for\ black\ holes,\ neutron\ stars\ are\ the\
smallest\ and\ densest\ currently\ known\ class\ of\ stellar\ objects\ \cite%
{ns1}, dense enough that\ one\ teaspoon\ (}$5\ mm$ {)\ of\ its\
material\ would\ have\ a\ mass\ over\ }$5.5\times 10^{12}\ kg$ {\ \cite%
{ns2}. Moreover the\ gravitational\ field\ of NS\ is\ about}$\ 2\times
10^{11}$ {\ times\ stronger\ than\ on\ Earth.To sum up the occurence
of EM radiations in NS and their strong gravitational effects makes it
suitable plasma environment for the validity of the present model.
Accordingly, for the numerical calculations we choose the typical plasma
parameters in the environment of neutron star} \cite{n5} : $n_{i0}\simeq
n_{e0}\simeq n_{0d}\sim 10^{27}cm^{-3},$ $T_{0,d}=(7\times 10^{5}-9\times
10^{5})$ $K$ and $B_{0}=10^{6}$ $G$. Inutility, the pre-requisite condition
to use DEMRP shown in Eq. (1) i.e., $\hbar \omega _{ps}(=)>>k_{B}T_{s}$ ($=$%
) is satisfied, then by utilizing these parameters along with surface
coordinate $x=10^{-4}$ $cm$ and $T_{0d}=7\times 10^{5}$ $K$ in (31) one can
easily calculate various physical parameters, such as the electric Alfv\'{e}%
n velocity $V_{E}(=E_{0}^{2}/4\pi \rho _{i}=4\pi \sigma _{se}^{2}\rho
_{i})=7.60\times 10^{6}$ $cm$ $s^{-1}$ and the gravitational radiation
acceleration $g_{d}^{\gamma }=$ $2.37\times 10^{11}$ $cm$ $s^{2}$. In order
to see gravity- radiation spectrum in dense plasma, let us substitute, $%
T_{0d}=7\times 10^{5}$, $n_{d}=1\times 10^{27}$ and $k=1\times 10^{6}$ in $%
\omega ^{2}=kg_{d}^{\gamma }$ to get $\omega =$ $4.87\times 10^{3}$ $Hz,$%
which belongs to the high frequency radio waves and the associated group
velocity of spectrum turns out to be $V_{gd}=243.5$ $cm/s$. {\ }%
Whereas for the REMRP, while using the data mentioned in \cite{n4} in Eq.
(2.43) of this paper\ i.e., $\omega ^{2}=k\gamma _{g}$, we get $\omega
=8.17\times 10^{4}Hz$ {.}

 {Generally speaking, the gravitational waves are the radiation
associated with the force of gravity, while in present analysis the pure
radiational gravity (RG) waves are shown to be function of wavelength
[Eq.(34)] and the associated inhomogeneous, effective radiative,
gravitational acceleration }$(g_{d}^{\gamma })$ {\ as function of
electron temperature is reported at dense plasma-vacuum interface}. {\
In other words the RG waves, calculated in the present work are independent
of gravitational acceleration }$\mathbf{g}$ {. To display a clear
vision of the impact of DEMRP on the propagation of RG waves and further to
compare our present findings with those of obtained in \cite{n4}, a
comparative plot is shown in Fig.1: the group velocity }$V_{gd}$  {%
[shown in Eq. (34)] verses}  {the scale length\ of pure RG waves, at
the surface of high density plasma and vacuum, \ for varying the temperature
of dense electrons, is displayed in fig.1(a), while the group velocity of RG
waves propagating at low density plasma -vacuum interface [obtained in Eq.
(2.44) of Ref. (20) and by using the data mentioned in this paper] is shown
in fig. 1 (b), for varying the temperature of rare (low density) electrons.
It may be noted here that since results displayed in Figs. 1(a) and 1(b) are
obtained in the presence of quite different radiation pressure i.e, }$\hbar
\omega _{ps}>>k_{B}T_{s}$ {\ and \ }$\hbar \omega _{ps}<<k_{B}T_{s}$%
 {\ respectively, so the parameters and hence the involved scale
lengths are quite different in both plots}.  {The comparative graphs
clearly admits that the group or phase velocity of RG depends on the choice
of radiation pressure i.e., dense or rare radiation pressures and hence on
the consequent plasma environments.}

 {Next, it is shown in (38) that the frictional term, appearing
through the Poynting vector, which involves the DEMRP, play a vital role to
\ enhance the surface oscillations of astrophysical objects via }$%
g_{d}^{\gamma ^{\prime }}$ {and }$\nu _{d}$. {\ The radiative
dissipative instability is plotted for }$T_{0,d}=(7\times 10^{5})$ {\ }%
$K$ {\ (red curve) and }$T_{0,d}=(9\times 10^{5}$ {\ }$K)$%
 {\ (blue curve) to see from the curves that the DEMRP or the
temperature of dense electrons play a vital role to increase the surface
perturbations. This might be an interesting result for the astrophysical
community to address\ the surface fragmentation of objects as function of
radiative, gravitational acceleration (}$g_{d}^{\gamma ^{\prime }}).$ {%
\ .}

A more clear comparison of present work with previous results achieved in
case of rare plasma \cite{n4} is presented in Table. 1, by calculating
various physical parameters such as: radiation pressures, electric Alfven
velocity, radiative gravitational acceleration, group velocity of
radiational gravity waves and growth rates of charged surface waves. For
numerical analysis, we choose typical plasma rare parameters mentioned in 
\cite{n4} and magnetic fusion plasma parameters \cite{13}, i.e., $%
n_{e0}\simeq (10^{13}-10^{14})cm^{-3},$ $T_{e0}=(10^{7}-10^{8})$ $K.$

\section{Tables and Figures}

$%
\begin{array}{ccccc}
Sr.No.\text{ } & 
\begin{array}{c}
Physical\text{ } \\ 
Parameter%
\end{array}
& 
\begin{array}{c}
Dense\text{ } \\ 
plasma%
\end{array}
& 
\begin{array}{c}
Rare\text{ } \\ 
plasma%
\end{array}
& 
\begin{array}{c}
Magnetic\text{ }fusion\text{ } \\ 
rare\text{ }plasma%
\end{array}
\\ 
1 & n_{e0}(cm^{-3}) & 1\times10^{27} & 1\times10^{12} & 1\times10^{14} \\ 
2 & T_{e0}(K) & 7\times10^{5} & 7\times10^{5} & 7\times10^{7} \\ 
3 & 
\begin{array}{c}
radiation\text{ } \\ 
pressure\text{ } \\ 
p_{e0,r}\text{ } \\ 
(dyne/cm^{2})%
\end{array}
& 
\begin{array}{c}
p_{d}^{\gamma}=\frac{U_{d}^{r}}{3}=\beta_{d}T_{d}^{5/2} \\ 
\beta_{d}=0.202(\hbar\omega_{pe})^{3/2}/(\hbar c)^{3} \\ 
=1.59\times10^{10}%
\end{array}
& 
\begin{array}{c}
p_{r}^{\gamma}=\frac{U_{r}^{r}}{3}=\alpha_{r}T_{r}^{4} \\ 
\alpha_{r}=\pi^{2}/45(\hbar c)^{3} \\ 
=2.01\times10^{8}%
\end{array}
& 
\begin{array}{c}
p_{r}^{\gamma}=\frac{U_{r}^{r}}{3}=\alpha_{r}T_{r}^{4} \\ 
\alpha_{r}=\pi^{2}/45(\hbar c)^{3} \\ 
=6.81\times10^{10}%
\end{array}
\\ 
4 & 
\begin{array}{c}
Electric\text{ }Alfven\text{ } \\ 
velocity\text{ }V_{E}(cms^{-1})%
\end{array}
& 7.60\times10^{6} & 7.60\times10^{6} & 7.60\times10^{7} \\ 
5 & 
\begin{array}{c}
Radiative\text{ } \\ 
gravitational\text{ } \\ 
acceleration\text{ } \\ 
(ms^{-2})%
\end{array}
& g_{d}^{\gamma}=2.37\times10^{11} & \gamma_{g}=6.68\times10^{7} & \gamma
_{g}=6.68\times10^{11} \\ 
6 & \omega\text{ }(Hz) & 
\begin{array}{c}
\omega=\sqrt{g_{d}^{\gamma}k} \\ 
=4.87\times10^{3}\text{ } \\ 
at\text{ }k=1\times10^{6}%
\end{array}
& 
\begin{array}{c}
\omega=\sqrt{\gamma_{g}k} \\ 
=8.17\times10^{4}\text{ } \\ 
at\text{ }k=1\times10^{2}%
\end{array}
& 
\begin{array}{c}
\omega=\sqrt{\gamma_{g}k} \\ 
=2.58\times10^{7}\text{ } \\ 
at\text{ }k=1\times10^{3}%
\end{array}
\\ 
7 & 
\begin{array}{c}
Radiative\text{ } \\ 
dissipative\text{ } \\ 
instability\text{ } \\ 
(s^{-1})%
\end{array}
& 3.33\times10^{6} & 2.30\times10^{12} & 4.97\times10^{14}%
\end{array}
$ \newline

Table 1: A comparison of different physical plasma parameters in dense and
rare plasma regimes.

\FRAME{dtbpFU}{6.5631in}{1.9865in}{0pt}{\Qcb{Group velocity of radiative
gravity waves in dense plasma $(V_{gd})$ [ by using Eq.(2.43)] is plotted
against the function of wave number $(k)$ by varying  {(a)} dense
electron temperature: $T_{0,d}=7\times 10^{5}$ $K$ (red curve) and $%
T_{0,d}=9\times 10^{5}$ $K$ (blue curve)at fixed values of $x=10^{-4}$ and
density $n_{0,d}=1\times 10^{27}cm^{-3}$. Subplot ( {b}) displays the
group velocity of radiative gravity waves in rare plasma $(V_{gr})$ [ by
using Eq.(2.44) of \protect\cite{n4}]against wave number $(k)$ by varying 
 {(b)} rare electron temperature: $T_{0,r}=7\times 10^{5}$ $K$ (red
curve) and $T_{0,r}=9\times 10^{5}$ $K$ (blue curve) at fixed density $%
n_{0}=1\times 10^{12}cm^{-3}$.}}{\Qlb{Fig. 1}}{Figure}{\special{language
"Scientific Word";type "GRAPHIC";maintain-aspect-ratio TRUE;display
"USEDEF";valid_file "T";width 6.5631in;height 1.9865in;depth
0pt;original-width 6.4999in;original-height 1.9476in;cropleft "0";croptop
"1";cropright "1";cropbottom "0";tempfilename
'S11BWA00.wmf';tempfile-properties "XPR";}}

\FRAME{dtbpFU}{4.8153in}{2.751in}{0pt}{\Qcb{The charged surface oscillations
dissipative gravitational radiation instability in dense plasma ($\hat{%
\protect\gamma}_{d}$) (as given by (2.38)) is plotted against the wavenumber 
$(k)$ for the same temperature range followd in Fig. (1).}}{\Qlb{Fig.2}}{%
Figure}{\special{language "Scientific Word";type
"GRAPHIC";maintain-aspect-ratio TRUE;display "USEDEF";valid_file "T";width
4.8153in;height 2.751in;depth 0pt;original-width 6.4999in;original-height
3.6979in;cropleft "0";croptop "1";cropright "1";cropbottom "0";tempfilename
'S11BWA01.wmf';tempfile-properties "XPR";}}

\section{Conclusions}

To summarize, one fluid magnetohydrodynamic approximation does not holds
good in case of charged surface waves at the interface of dense radiative
electron-ion plasma under the influence of thermal and radiation pressures.
An important aspect of the present findings is that the radiative
gravitational acceleration becomes function of the electron temperature. It
is shown that gravitational radiation acceleration plays a vital role to
stabilize the surface charge instability. For a particular condition, it is
focused that pure, inhomogeneous (as a function of electron temperature)
gravitational- radiation waves, of typical frequencies\ associated with DRP
may propagate on the interface as a function of wavelength. It is focused
that in case of high density and low temperature astrophysical plasma, the
frequency of the obtained radiative gravitational waves is found to lie in\
the range of high frequency radio waves, while in case of rare laboratory
plasma, the frequency of these waves is found to lie\ within the very low
frequency radio waves of the electromagnetic radiation spectrum. Further, it
is depicted that if radiations acts for a short time, dissipative
instability arises as a function of gravitational radiation acceleration.
The growth rate of frictional instability is found to be significant
function of electron temperature.

\end{document}